 \documentstyle[12pt]{article}
\newcommand{\be}{\begin{equation}}
\newcommand{\ee}{\end{equation}}
\newcommand{\bea}{\begin{eqnarray}}
\newcommand{\eea}{\end{eqnarray}}

\begin{document}
 \begin{titlepage}

\begin{flushright}
CERN-TH.7240/94
\end{flushright}
\vspace{20 mm}

\begin{center}
{\huge The Scattering of Strings}

\vspace{5mm}

{\huge  in a }

\vspace{5mm}

{\huge Black-Hole Background}

\end{center}

\vspace{10 mm}

\begin{center}
Ulf H. Danielsson\\
Theory Division, CERN, CH-1211 Geneva 23, Switzerland

\end{center}

\vspace{2cm}

\begin{center}
{\large Abstract}
\end{center}

In this paper, correlation functions of tachyons in the
two-dimensional black-hole background are studied.
The results
are shown to be consistent
with the prediction of the deformed matrix
model up to external leg factors.

\vspace{3cm}
\begin{flushleft}
CERN-TH.7240/94 \\
May 1994
\end{flushleft}
\end{titlepage}
\newpage

\section{INTRODUCTION}

In this letter, I will give some results on tachyon scattering amplitudes
in the background of
a two-dimensional black hole \cite{wad,wit1}. For this purpose,
I will
use the techniques developed in \cite{igor} based on Ward identities.
The Ward identities will provide powerful constraints on the
correlation functions that are otherwise so hard to calculate.
The black hole will be constructed as a perturbation  around the
ordinary linear dilaton vacuum. Indeed, in \cite{ber} it was shown
that the action of the $SL(2,R)/U(1)$
coset model \cite{wit1}, which describes a two-dimensional black hole,
can be written as
\be
\frac{1}{2\pi} \int(\partial X
\bar{\partial}X +\partial \phi \bar{\partial}\phi -
\sqrt{2} R \phi + M{\cal B} )    ,
\ee
where
\be
{\cal B} \sim
(3i \partial X +\partial \phi )
(3i \bar{\partial} X +\bar{\partial} \phi ) e^{-2\sqrt{2} \phi }
\label{svsc}
\ee
is the `black-hole screener'. For further discussion, see
\cite{eguchi2}. In the brave
spirit of \cite{goul}, I will consider tachyon correlation functions
in the presence of an integer number of such black-hole screeners.
The hope is that this will teach us something about the true
tachyon correlation functions in the black-hole background.
However, this na\"\i ve
perturbative approach has severe limitations.
This will be apparent in the form of the leg factors.
I will return to this important issue
in the last section.

The main purpose of this letter is to show that the correlation
functions of the deformed matrix model [7--13] are, in a highly
non-trivial way, consistent with the black-hole Ward identities.
More details will be given in a future publication.

\section{WARD IDENTITIES IN TWO-DIMENSIONAL STRING THEORY}

\subsection{Without Black-Hole Screeners}

In \cite{igor} Ward identities were derived, which related tachyon
scattering amplitudes with different numbers of tachyons. These
Ward identities were
shown to produce recursion relations that determined
all the amplitudes. Compared to more standard calculations, see
e.g. \cite{fran}, this was a remarkable simplification. Below I will
briefly summarize the results of \cite{igor}.

The Ward identities are derived by using the charges
\be
Q_{m,-m} =
\oint \frac{dz}{2\pi i} W_{J,m}
-\oint \frac{d\bar{z}}{2\pi i} c \Psi _{m+1,-m} \bar{X}_{m,-m}  ,
\label{ladd}
\ee
where
\be
W_{J,m}
 =\Psi _{J+1,m}(z) \bar{\cal O} _{J,m} (\bar{z})   ,
\ee
first constructed by Witten in \cite{wit2}. The fields $\Psi$
correspond to the gravitationally dressed primary fields
\be
\Psi _{J,m}
= \psi _{J,m} (z) e^{\sqrt{2}(-1+J) \phi (z)}   .
\ee
The ${\cal O}$'s are elements of the  ground ring.
The $X$-fields have ghost number $-1$, but their precise
form need not concern us.

In \cite{igor} the action of the currents on tachyons was calculated.
In terms of the rescaled tachyons, $\tilde{T}$, where
\be
T_{p}^{\pm}  = \pi \Delta (1\mp \sqrt{2}p) \tilde{T}_{p}^{\pm}
\ee
($\Delta (x) = \Gamma (x)/\Gamma (1-x)$), it was found that
\be
W_{m,-m} (z) c\bar{c} \tilde{T}^{-}_{-p} (0) =
\frac{1}{z} c \bar{c}
(\sqrt{2}p +2m) \tilde{T}^{-}_{-p-m\sqrt{2}} (0) +...
\ee
$\pm$ refers to chirality. A positive-momentum rightly dressed
tachyon is defined to have positive chirality.
I will follow \cite{igor} and write this, in an obvious notation,
as
\be
Q_{m,-m} | \tilde{T}_{-p}^{-} \rangle =(\sqrt{2}p+2m)
| \tilde{T}^{-}_{-p-m\sqrt{2}}\rangle       .
\label{neg}
\ee
As is clear from above, the first term
in (\ref{ladd}) has ghost numbers $(0,0)$, while the second
term has
$(1,-1)$. Since the tachyon (if fixed) has ghost numbers $(1,1)$,
and has no $(2,0)$ part, the second term
cannot contribute to the right-hand side
(apart from a BRST-exact state). It is only for rightly
dressed non-tachyonic states that one gets a non-trivial
contribution. These are the new moduli discussed in \cite{wit3}.

The same current acting on tachyons of the opposite chirality gives
a very different result.
In general
\be
Q_{m,-m} | \tilde{T} ^{+}_{p_{1}},...,\tilde{T}^{+}_{p_{2m+1}}\rangle =
(2m+1)! \left(\sqrt{2} \sum _{i=1}^{2m+1}
p_{i} -2m\right)
| \tilde{T} ^{+}_{\sum _{i=1}^{2m+1} p_{i} -\sqrt{2}
m}\rangle         .
\ee
It is hence
capable of changing the number of tachyons. This is the
reason why it is possible to construct powerful recursion relations.
For instance, an insertion of the $W_{1/2,-1/2}$
current leads to the relation
\be
(N-2)A(N) = (N-2)(N-3) A(N-1)
\ee
among the tachyon amplitudes $A(N)$, which have $1$ negative and
$N-1$ positive chirality tachyons.
This, then, implies the well-known result
\be
A(N) = (N-3)!     ,
\ee
given that $A(3)=1$.
Note that as compared to \cite{igor}, I have rescaled also the
negative chirality tachyon. This would really give zero
since it sits on discrete momenta
(the bulk amplitudes of these renormalized tachyons are zero),
but the zero is compensated by including the volume factor leading
to a finite result \cite{davig}.
This is the convention of collective field
theory.

\subsection{With Black-Hole Screeners}

Let me now generalize the above construction to correlation functions
involving black-hole screeners. I will
denote such correlation functions
by $A(N,m)$, where $N$ is the total (including the negative chirality
one) number of tachyons and $m$ is the number of black-hole screeners.
Momentum conservation implies
\be
p_{N} =-\frac{N+2m-2}{\sqrt{2}}    .
\ee
I will use the charges $Q_{n/2,-n/2}$, with $n \geq 2m-1$, to derive the
Ward identities. From momentum conservation it follows that
$Q_{n/2,-n/2}$ transform $n+1-2k$ positive chirality tachyons and
$k$ screeners into one tachyon. I will assume, as an ansatz, that
the precise form is
$$
Q_{n/2, -n/2} \mid \tilde{T}_{p_{1}}^{+}... \tilde{T}^{+}
_{p_{n+1-2k}}{\cal B}^k \rangle
$$
\be
= 2^{-k} k! (n+1-2k)!
\left(\sqrt{2} a_{n,k} \sum _{i=1}^{n+1-2k} p_{i} +b_{n,k}\right)
\mid \tilde{T} ^{+}
_{\sum _{i=1}^{n+1-2k} p_{i} -n/\sqrt{2}} \rangle    .
\label{pos}
\ee
The $a_{n,k}$ and $b_{n,k}$ are
coefficients depending on $n$ and $k$ only.
The prefactor is for latter
convenience. This ansatz is based on the assumption of
factorization into leg factors. If this assumption is true,
it is clear that factors of $\Delta$ must appear,
in the same way as before when the black-hole screeners are
involved. Furthermore, the residual dependence on momenta is
clearly symmetric in the different momenta, but must also be at most
linear in order for the resulting recursion
relations, for all $N$, not to
depend on individual momenta when all contributions are
summed up. This is a prerequisite for factorization into
leg factors. For each black-hole screener
there will appear a regulated zero
$\sim 1/\log M$ (see \cite{fran} for a discussion on zeroes and poles
of correlation functions with dicrete states),
which, symbolically, I will write as $\Delta (1)$.
The $1/\log M$ will be absorbed into ${\cal B}$, or rather
$M$. This is no different from standard $c=1$ where
$\Delta/\log \Delta \rightarrow \mu$.

Equations (\ref{neg}) and (\ref{pos})
are the only possible non-trivial
contributions. This can be seen by using momentum conservation and
examining the singularities of the contractions.
All other possibilities give at best discrete states at non-discrete
momenta. For $n < 2m-1$, however, discrete states at discrete momenta
would appear, giving more complicated Ward identities.

It is straightforward to write down the generalization of the
$c=1$ Ward identities using the above results.
It follows that
$$
(N+2m-2)A(N,m) = (N-2)...(N-n-1)(N+2(m-1)(n+1) +n) A(N-n,m)
$$
$$
+
\sum _{k=1}^{m} m(m-1)...(m-k+1)
(N-2)...(N-n-1+2k)
$$
\be
\times
(a _{n,k}(n+1-2k)(N+2m-2) +b_{n,k} (N-1)) A(N-n+2k,m-k) ,\label{ward}
\ee
when all possible non-vanishing cancelled propagators are taken into
account.
The $a _{n,k}$ and $b _{n,k}$
could in principle be calculated directly
using the methods of \cite{igor}. I will, however, not attempt this
in this paper since a careful study of regularization is first needed.

Let me repeat the logic of the present discussion. {\it If} the
correlation functions factorize as in standard $c=1$, {\it then}
they must satisfy the Ward identities (\ref{ward}) for
some values of $a_{n,k}$ and $b_{n,k}$.
I stress that
even without explicit values for the coefficients
$a_{n,k}$ and $b_{n,k}$
these are very powerful Ward identities.
Below I will show that the deformed matrix model results
are compatible with (\ref{ward}), while some other proposals are not.
But first I need to explain the prediction of the deformed matrix model.
Remember that this model has been argued [7--13]
to be the matrix model realization of the black hole.

\section{SOLUTIONS OF THE WARD IDENTITIES}

\subsection{The Deformed Matrix Model Prediction}

The deformed matrix model is obtained by adding a piece
$M/2x^{2}$ to the matrix model potential. The potential becomes
\be
-\frac{1}{2\alpha '} x^{2} +\frac{M}{2x^{2}}  ;
\ee
$M$  will be positive. The position of the Fermi level
is, as usual, measured in terms of its deviation from zero, i.e. $\mu$.
However, it is now
possible to define a double scaling limit, even when $\mu =0$. One then
needs to keep
$\hbar /M^{1/2}$ fixed,
which will be the string coupling constant.

Special cases of tachyon correlation functions have been calculated
in several papers, [7--9,11--13]. In \cite{dual} the general formula
(in the case with $N-1$ tachyons of the same chirality)
was given up to genus one. The genus-zero piece, at $\alpha ' =2$,
is
\be
\langle  \tilde{T}^{+}_{p_{1}}...\tilde{T}
^{+}_{p_{N-1}} \tilde{T}^{+}_{-p} \rangle
=
(N-3)!!
(\sqrt{2}p-2)(\sqrt{2}p-4)
...(\sqrt{2}p-(N-4)) M^{p/\sqrt{2} -N/2 +1}   ,
\label{tmcorr}
\ee
when normalized to collective field theory.
This result is valid only for even $N$: for odd $N$ the matrix model
gives zero.
Put $p=\frac{N+2m-2}{\sqrt{2}}$ and take $m$ derivatives of
(\ref{tmcorr}) to get
\be
A(N,m)
=2^{-m} (N-3)!! (N+2m -4)!!    .
\label{defrel}
\ee
This, then, is the deformed matrix model suggestion for the
correlation functions with black hole screeners.

\subsection{A Check of the Ward Identities}

Let us now check whether the matrix model prediction can be made to
satisfy the Ward identity  (\ref{ward}). As I will discuss in the
last section, the vanishing
of the matrix model odd-point functions is
a consequence of a difference in the leg factors coming from a
different choice of vacuum. It is natural to assume that when the
perturbative vacuum is used, the odd-point functions are given by
a direct `analytic continuation' of the even-point result. This
I will use below. In the
next subsection I will provide some evidence for this by showing that
it is true for the three-point function.

It is straightforward
to substitute (\ref{defrel}) into the Ward identities and
determine the necessary $a_{n,k}$ and
$b_{n,k}$ one by one in $k$; $m=1$ provides $k=1$, $m=2$ provides
$k=2$ (given the $k=1$), etc. The answer is
\be
a _{n,k} = \frac{n}{k!} (n-2)(n-4)...(n-2k+2)
\ee
and
\be
b_{n,k} = -n a_{n,k}  .
\ee
I stress that
{\it it is a highly non-trivial property of the deformed
matrix model correlation functions that coefficients can be found
such that the Ward identities are obeyed.  }
For instance, it can be checked that
other proposals, such as
$A(N,m) = (N+2m-3)!\frac{m!}{(2m)!}$ (which
would correspond to $c=1$ correlation functions with
$\mu ^2 \rightarrow M$) fail the test at $m=2$.

\subsection{Some Explicit Examples}

Even if a direct calculation of the scattering amplitudes is in general
very difficult (which is one reason to consider Ward identities),
there are some cases where the calculation can
be done and hence used as a check of the above results. This is
so when $m=1$.
A convenient representation of the black-hole
screener is
\be
\frac{1}{ia} \lim _{y \rightarrow z} \partial _{y}
:e^{-iaX(z) +b\phi (z) +iaX(y) +c\phi (y)}:
\ee
where $b+c=-2\sqrt{2}$, and the same for the anti-holomorphic piece.
With $a=6\sqrt{2}$ and $b=0$ eq. (\ref{svsc}) is
reproduced. I will, for the moment, keep
the polarization tensor, i.e. $a$ and $c$, arbitrary.
This can be taken as a useful check on the calculation. As explained
in \cite{pol}, one of the graviton polarizations is pure gauge, i.e.
it can be written as $L_{1}$ of something. In general, by using the
gauge freedom, any special state can be written in a form where it
contains no $\partial ^{n} \phi$. Hence, the final answer should
not depend on the polarization.
However, the resulting integrals simplify only for
certain peculiar values of the momenta.
Let me pick $p=\frac{N}{\sqrt{2}}$,
$p_{1}=...=p_{N-2}=\frac{\sqrt{2}c}{c-a}$, and consequently
$p_{N-1}=\frac{N}{\sqrt{2}} - (N-2)\frac{\sqrt{2}c}{c-a}$.
The integral to be performed, with the black-hole screener at $\infty$, is
\be
-\frac{1}{a^{2}}\left(
(a+c)\frac{N}{\sqrt{2}} -\sqrt{2} c\right) ^{2}
\prod _{i=1}^{N-2} \int d^{2} z_{i} \mid
z_{i}\mid ^{2\alpha}
\mid 1-z_{i} \mid ^{2\beta} \prod _{i>j}^{N-2} \mid
z_{i} -z_{j}\mid ^{2\gamma}        ,
\ee
where
\be
\alpha = \left( 1-\frac{2c}{c-a}\right)
N +\frac{2c}{c-a} -2   ,
\ee
\be
\beta = N-2 -(N-3)\frac{2c}{c-a}
\ee
and
\be
\gamma = \frac{4c}{c-a} -2               .
\ee
The integral can be evaluated by using the results of \cite{dots}
and the result is, with the volume factor,
\be
2(N-2)!\Delta (1-\sqrt{2}p )^{N-2}
\Delta (1-\sqrt{2}p_{N-1}) \Delta (1-N) \Delta (1)  ,
\label{tjohej}
\ee
with no dependence left on the polarization tensor as promised.
We recognize the familiar leg factors and see that the remaining
non-factorizable piece agrees with (\ref{defrel}) at $m=1$ up to
the normalization of ${\cal B}$.

Let me now consider the three-point in a little more detail.
According to (\ref{defrel}) it is given by
\be
2^{-m} \frac{(2m-1)!!}{m!} M^{m}    .
\label{a3m}
\ee
It is now necessary to
comment on \cite{beck}, where the three point is calculated
for general $m$ using continuum methods.
It was found that the three-point is the same as for $c=1$ up to a factor
\be
\frac{1}{\Gamma (\frac{x+4}{8})}
\frac{\Gamma (m+\frac{x+4}{8})}{\Gamma (m+1)}  ,
\label{norm}
\ee
where $x$ is an unknown parameter introduced by the
regularization. In \cite{beck} $x$ was fixed so that the factor
(\ref{norm}) was one. However, one can argue in favour of another
prescription. Let me demand that $m$ and the Liouville momentum of
the negative chirality tachyon are untouched by the regularization.
After all, it is only for integer $m$ that we can perform the
calculation.
As is clear from \cite{beck}, this means that $x=0$.
The three-point then becomes precisely (\ref{a3m}).
This is the only way to remain consistent with (\ref{tjohej}),
where this regularization prescription has been used implicitly.
One can also check that the deformed matrix model
expression for the two-point, $\frac{1}{2m} M^{m}$, is
in agreement with the continuum calculation.

\section{CONCLUSIONS}

In this paper we have seen how the black-hole tachyon correlation
functions compare
with those of the deformed matrix model. We have seen that the
latter obey highly non-trivial Ward identities derived for
the black-hole background. I have also provided some examples of
explicit calculations, where the agreement can be verified.
There are therefore  reasons to believe that
the deformed matrix model really {\it is} a black hole.
It should be possible to complete a rigorous proof
along the lines of this paper.

A difference between the results is the form of the external leg
factors or rather the position of the poles in these factors.
After all, it is only the latter that are universal
and can be confidently predicted
by the matrix model without further physical input. In the continuum
calculation we are implicitly using the same vacuum as in the
linear dilaton theory and we are bound to obtain the same external leg
factors in our perturbative treatment. It is easy to see, however,
that this vacuum is an unfortunate choice. In the black hole
we often demand periodicity in Euclidean time. This is just a
choice, but a sensible one. It means that we have chosen a
vacuum corresponding to an equilibrium eternal black hole at the
Hawking temperature.
The particular value of the Euclidean compactification radius
$R$, i.e. the inverse temperature, is obtained if one
insists on no conical singularity at the origin of Euclidean space,
i.e. the horizon.
In the two-dimensional black hole, the compactification radius differs
between the semiclassical case \cite{wad,wit1}
and the `exact' metric of \cite{dvv},
being $1/\sqrt{2}$ and $3/\sqrt{2}$,
respectively. Recall that I use conventions such that
$\alpha ' =2$. For comparison, the self dual
radius is $R _{S} =\sqrt{2}$ and the Kosterlitz--Thouless phase
transition takes place at $R_{T} =2\sqrt{2}$. (Since the above values
are below $R_{T}$, this might be a problem.)
In standard $c=1$ the poles occur at $p =\frac{n}{\sqrt{2}}$ for
all integers $n$. If Euclidean time is compactified the allowed
momenta must be of the form $\frac{n}{R}$. Clearly all states
are allowed at $R=R_{S}$. The situation is different when we pick
one of the black-hole radia suggested above. In both cases the odd
poles are not allowed! This is the reason why we get double spacing
of correlation function poles. The matrix model is clever enough
to spot this problem. In \cite{grop} it was shown that matrix eigenvalue
wave functions that might give rise to poles other than the ones
above were in general non-normalizable. The perturbative continuum
calculation, however, is too crude to take this into account.

To conclude, there is strong evidence that the deformed matrix model
is describing a two-dimensional black hole. If this is indeed true,
we have a powerful tool at our disposal to explore
stringy quantum black-hole phenomena.

\section*{Acknowledgements}

I would like to thank K. Becker, M. Becker and, especially,
R. Brustein for discussions.

\end{document}